\documentclass[11pt,prb,reprint,superscriptaddress,nobibnotes,nohyperlinks]{revtex4-1}

\usepackage[hidelinks,breaklinks=true]{hyperref}
\usepackage{graphicx,amsmath,amssymb,color}
\usepackage{tabularx, ctable}
\usepackage{ulem}
\usepackage{bm}
\usepackage{siunitx}
\usepackage{float}
\usepackage{gensymb}
\usepackage{wrapfig}
\usepackage{dirtytalk}
\usepackage{xcolor}
\usepackage{xpatch}
\usepackage{upgreek}
\usepackage[T1]{fontenc}

\makeatother

\begin{document}

\title{Surface defects in carbon-doped hexagonal boron nitride for negative-contrast direct laser writing}

\author{Dmitrii Litvinov}
\affiliation{Department of Materials Science and Engineering, National University of Singapore, 117575 Singapore}
\affiliation{Institute for Functional Intelligent
Materials, National University of Singapore, Singapore, 117575, Singapore}

\author{Virgil Gavriliuc}
\affiliation{Institute for Functional Intelligent
Materials, National University of Singapore, Singapore, 117575, Singapore}

\author{Magdalena Grzeszczyk}
\affiliation{Institute for Functional Intelligent
Materials, National University of Singapore, Singapore, 117575, Singapore}

\author{Kristina Vaklinova}
\affiliation{Institute for Functional Intelligent
Materials, National University of Singapore, Singapore, 117575, Singapore}

\author{Kenji~Watanabe}
\affiliation{Research Center for Functional Materials, National Institute of Material Science, Tsukuba 305-0044, Japan}

\author{Takashi~Taniguchi}
\affiliation{International Center for Materials Nanoarchitectonics, National Institute of Material Science, Tsukuba 305-0044, Japan}

\author{Kostya~S. Novoselov}
\affiliation{Department of Materials Science and Engineering, National University of Singapore, 117575 Singapore}
\affiliation{Institute for Functional Intelligent
Materials, National University of Singapore, Singapore, 117575, Singapore}

\author{Maciej Koperski}
\email{Correspondence to: msemaci@nus.edu.sg}
\affiliation{Department of Materials Science and Engineering, National University of Singapore, 117575 Singapore}
\affiliation{Institute for Functional Intelligent
Materials, National University of Singapore, Singapore, 117575, Singapore}


\begin{abstract}
\textbf{
Radiative defects in hexagonal boron nitride (hBN) are active in a broad spectral range from deep ultraviolet to near-infrared wavelengths. Representatives of these defects act as bright single photon sources, spin-1 systems, and multiproperty atomic-scale sensors. They are predominantly investigated in bulk hBN films, where defects are decoupled from surface and interfacial effects. Here, we demonstrate a novel class of surface defects optically active in the green/yellow visible spectral range, which exhibit photophysical properties distinct from their bulk counterparts. High-power resonant laser illumination quenched the emission from the ensemble of such defects, which was attributed to a light-driven structural reconfiguration. The quenched defects were found to recover their emissive capabilities via a thermal cycling process, revealing an activation energy of 24.5~meV for the structural transition. Alternatively, permanent quenching of the defects was triggered by surface chemistry, involving lithiation-enabled attachment of functional groups. These mechanisms were utilized to realize negative-contrast direct laser writing, designing arbitrary geometric emissive patterns on demand in a microscopic configuration. The surface-active radiative centers in hBN appear particularly attractive for exploring environmental sensitivity, surface science, and coupling to photonic structures or electronic devices by taking unique advantage of the two-dimensional characteristics of the host lattice.
}

\end{abstract}

\maketitle

Defect centers in hexagonal boron nitride (hBN) are characterized by diverse atomic and electronic structures, leading to a library of optically active defects differentiated by distinct optoelectronic properties \cite{hBN_SPE_powder}. The interest in creating and characterizing defects is driven predominantly by their potential applications in three areas \cite{2D_photonics_roadmap}: 1) single photon emission for quantum telecommunication \cite{SPE_2D_review}, 2) coherent spin manipulation for optical qubit platforms\cite{gottscholl2021room,stern2024quantum,gao2025single}, and 3) atomic-scale sensing of multiple physical quantities, including magnetic \cite{healey2023quantum,huang2022wide,das2024quantum} or electric fields \cite{noh2018stark,xia2019room}, temperature\cite{liu2025temperature}, or pressure \cite{gottscholl2021spin,zeng2025ambient}. All these applications depend on the positioning of defects with respect to the surface of the host lattice. Deeply buried defects in bulk films can be decoupled from the environmental effects. These can include charge fluctuations that cause instabilities in the spectral characteristics and emission intensity of single photon sources \cite{chen2016characterization,ha2015size,li2022role}, or constitute decoherence channels for optically driven spin centers \cite{sangtawesin2019origins,kim2015decoherence}. Defects in thin films are more susceptible to the influence of strain fields and modulations to the dielectric environment, which lead to the emergence of inhomogeneous broadening effects of the radiative resonances \cite{hBN_dielectric_sensitivity}. On the other hand, defects close to the surface can be coupled more efficiently to photonic structures and electronic devices \cite{hBN_LED, quantum_LED_perspective}, while exhibiting enhanced sensitivity to both intrinsic and extrinsic physical quantities for sensing applications \cite{sensing_arxiv}. Two-dimensional characteristics of the host crystal lattice are particularly beneficial for managing the challenges and benefits arising from defects being positioned at varied distances from the surface. This originates from a large surface-to-volume ratio and the ability to fabricate films of monolayer thickness, which exhibit sub-nm thickness for hBN crystals.

Despite these advantages of van der Waals systems, defect centers in hBN are predominantly investigated in bulk hBN layers, while the demonstrations of light emission from defects in atomically thin layers are quite rare. Commonly used fabrication procedures of defects in bulk crystals with reproducible spectral characteristics, such as ion irradiation \cite{liang2025site, liang2023high}, annealing in dopant-rich atmospheres \cite{koperski2020midgap}, or \textit{in-situ} doping arising from the presence of the carbon source within the reactor environment \cite{mendelson2021identifying, krivobok2022tin, hBN_MOVPE} typically implant the defect centers far away from the surface. The penetration depth of light ion beams and diffusion-driven dopant profiles can extend even beyond the micrometer length scale \cite{kianinia2020generation}, while the fabrication of mono- and multilayer hBN films hosting radiative centers from the defect-enriched bulk crystals is technologically challenging \cite{monolayer_hBN_identification}. Here, we demonstrate the existence of radiative defects residing on the basal plane of bulk layers of carbon-doped hBN, providing an alternative approach to isolating surface states from the bulk, circumventing the fabrication of atomically thin films. The unique characteristic of the surface defects is their controllable and reversible quenching by resonant microscopic laser excitation. The recovery process through thermal cycling reveals a 24.5 meV energy barrier characterizing the transition between radiative and non-radiative states, indicating a light-driven structural reconfiguration as a plausible mechanism. The permanent quenching was achieved by changing the chemical composition of the defects by the covalent attachment of functional groups via classical surface chemistry, which constitutes a proof of the pure surface nature of this class of defects.

The reversible laser quenching of the defects was utilized to create arbitrary geometric patterns, demonstrating microscopic laser writing capabilities. Creating resettable negative contrast patterns in the microscale provides novel non-invasive methods of spatial patterning of emissive films. The surface nature of the defects differentiates them from the bulk defects, proving alternative approaches for the development of environmental sensing, interfacial coupling, and local manipulation via methods of surface chemistry and physics.

\textbf{Results.}

In this work, we study the optical properties of carbon-doped hBN flakes mechanically exfoliated from a bulk crystal, which were synthesized by the high-pressure high-temperature method. The carbon doping was achieved by post-growth annealing in a graphite furnace. The previously reported secondary ion mas spectroscopy of those crystals revealed volumetric density of carbon impurities to reach $10^{20}$~cm$^{-3}$ at the surface of the crystals with a decreasing gradient down to $10^{18}$~cm$^{-3}$ occurring at a length scale of about $1~\upmu$m \cite{hBN_carbon_doping}. The scanning tunnelling microscopy of devices comprising a graphite substrate and few-layer carbon-doped hBN films demonstrated a large population of carbon substitutions for boron, nitrogen vacancies, and combinations of vacancies and impurities forming multi-site defects \cite{Qiu2024}. The mechanism of stabilization of complex defects was attributed to the formation of molecular-like states embedded within the band gap of hBN when donor-like and acceptor-like defects are combined, which lowers the formation energy of complex multi-site defects \cite{carbon_doped_hBN_DFT, hBN_ab_initio}. Consequently, high-symmetry defect structures, such as carbon hexagonal rings embedded within hBN layers, constitute energetically favorable structures. 

\textbf{Photoluminescence spectrum}

An example of a low-temperature (4~K) photoluminescence (PL) spectrum of exfoliated bulk carbon-doped hBN film under sub-bandgap laser excitation is presented in figure 1(a). The spectrum is spatially averaged over the area of the hBN film, based on the hyperspectral mapping in a microscopic configuration. Under such excitation conditions, the PL signal can be attributed to the ensemble of radiative defect centers. We identified three dominant features related to the emission of distinct types of defects marked as resonances A, B, and C in figure 1(a). The defects B and C were investigated previously \cite{koperski2020midgap}, and attributed to a carbon dimer coupled to a vacancy (B) and a nitrogen vacancy (C) \cite{hBN_dielectric_sensitivity}. Defect A was found to dominate the PL response for a specific batch of carbon-doped hBN crystals, which can be linked to a conjunctive effect of intrinsic carbon doping and environmental exposure that modifies the surface of the exfoliated films. The carbon diffusion into hBN creates a concentration gradient of the substitutional atoms, with the exact depth profiles highly sensitive to process parameters (e.g., annealing temperature or time) and intrinsic defect density. Moreover, the reduced formation energies of carbon-carbon and carbon-vacancy defects are responsible for the stabilization of increasingly more complex defect structures. The dilution of the surface concentration of carbon via interlayer diffusion, occurring synergistically with the combinatorial merging of intentionally introduced carbon substitutions and intrinsic vacancies, constitutes an inherently stochastic process. Consequently, the ratio of surface-to-bulk defects, which influence the chemical reactivity of the surface of the crystals with the environment, exhibits variations between crystals resulting from precisely controlled growth and annealing processes. This interpretation aligns with the results of the comparative characterization of the properties of different defects, distinguishing the bulk defects (B, C) from surface defects (A).

The spatial distribution of the radiative defects constitutes one of the differentiating factors between defects A, and B/C. An atomic force microscopy image of an example carbon-doped hBN flake is demonstrated in figure 1(b). The cross-sectional line profile in figure 1(c) shows that the film is characterized by 55~nm thickness. The microscopic mapping of the PL response revealed that the distribution of the signal from defect A is sensitive to the topography features, as shown in figure 1(d), revealing enhanced PL intensity along the edges of the film, thickness steps, or interfacial features such as wrinkles. The distribution of the PL signal from defects B and C is significantly more homogeneous, as demonstrated in figures 1(e,f).

\begin{figure*}[t]
    \centering
    \includegraphics[width=\textwidth]{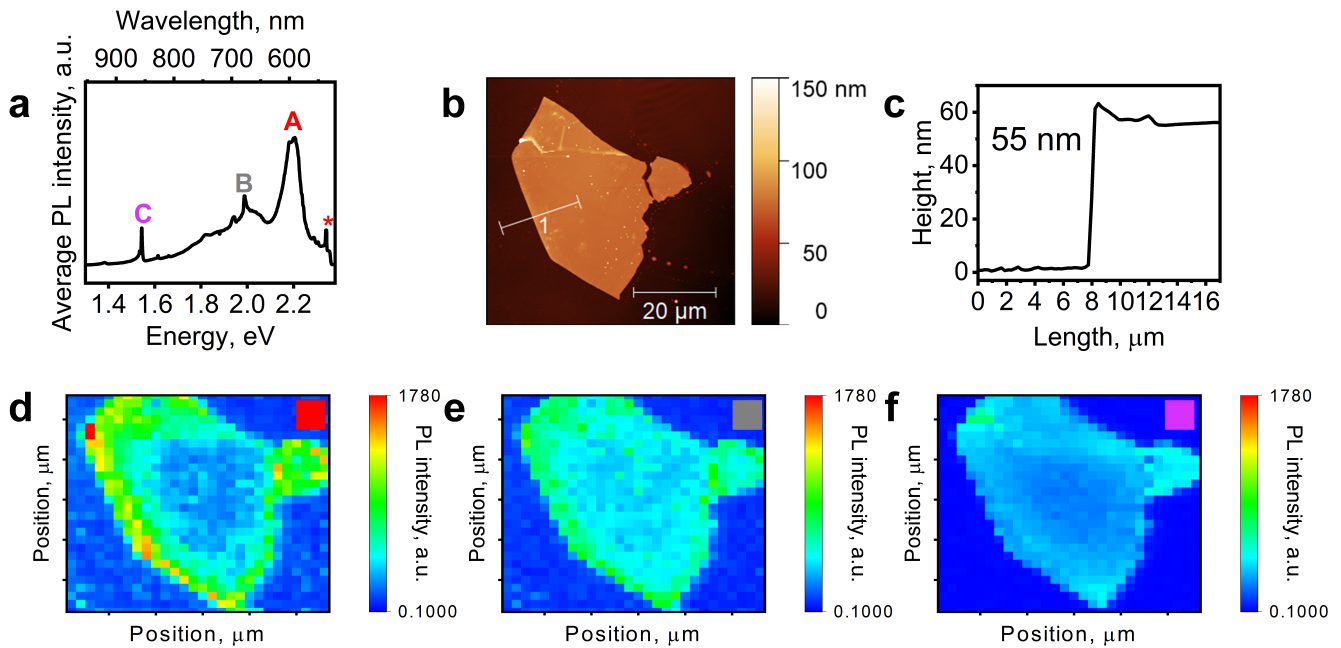}
    \caption{Optical response and structural characterization of carbon-doped hBN films. (a) The low temperature (4~K) photoluminescence (PL) spectrum of a mechanically exfoliated carbon-doped hBN film. The spectrum is spatially averaged over the entire area of the hBN flake based on raster mapping of the PL spectra in a microscopic configuration. The red star indicates the Raman scattering resonance related to the optical phonon from the silicon substrate. A, B, and C mark the positions of zero-phonon lines for different radiative defect centers. (b) An atomic force microscopy image of a carbon-doped hBN flake. (c) A cross-sectional height profile along line 1 from panel (b). (d-f) - $\upmu$PL maps of the carbon-doped hBN flake obtained by monitoring the integrated PL intensity of defects A, B, and C, respectively. The intensity color scale is logarithmic.}
    \label{fig1}
\end{figure*}

The PL spectrum of defect A from an edge location with enhanced signal intensity is presented in figure 2(a) in an energy scale relative to the highest energy resonance corresponding to the zero-phonon line (ZPL). Two replicated resonances are observed at the energy of 173 and 346~meV, which are in agreement with one- and two-phonon replicas related to the optical phonon in hBN lattice. This demonstrates that the electrons participating in the optical transitions of defect A are coupled to the vibrational motion of the host lattice within the one-dimensional Franck-Condon model of intradefect excitations with a Huang-Rhys factor of $1.17 \pm0.06$. Phonon-replicated absorption resonances are also observable in the photoluminescence excitation (PLE) spectra presented in figure 2(b). Three quasi-absorption resonances correspond to the direct intradefect absorption associated with ZPL (not fully captured due to limitations in spectral filtering), and two other excitation mechanisms through concurrent creation of phonons. Even though the main resonance is not fully observed, we estimated a Stokes shift to be $\mathrm{E_{Stokes}} = 33.1 \pm 0.7$ meV based on the energy difference between one-phonon replicas observed in the PL and PLE spectra (see figure 2(b)). That is a typical value for defects in hBN characterized by a finite electron-phonon coupling strength quantified by a Huang-Rhys factor comparable to defect A.

\textbf{Quenching and recovering of the photoluminescence signal.}

A unique characteristic of defect A is the systematic quenching of the PL intensity under high-power laser illumination and recovery via thermal cycling, as illustrated in figure 2(c). The experiment was conducted using multiple consecutive laser irradiation protocols to read out and modify the radiative states. The read-out process was realized via a PL mapping experiment at 4~K under 515 nm diffraction-limited microscopic laser excitation with a power of 200 $\upmu$W delivered to the sample and 1 s of exposure time per point. The spatially averaged PL spectrum from the entire area of the flake (black curve in figure 2(c)) represents the initial radiative state of the sample. A second irradiation process (quenching) was conducted with the increased laser power of 10 mW and increased exposure time of 60 s. The following read-out of the radiative state of the sample revealed a reduction of the total PL intensity from defect A by a factor of 15, as shown by the blue curve in figure 2(c). A thermal cycling procedure partially reversed the quenching. The sample was warmed up to 300 K, kept at that temperature for 30 min, and cooled back down to the base temperature of 4~K. The final read-out of the emissive state (red curve in figure 2(c)) demonstrated an increase of the total PL intensity by a factor of 7.5 with respect to the quenched state, which constitutes 50 \% of the PL intensity recovery from the initial state. No changes in the PL intensity from bulk defects were observed during this procedure, as illustrated for defect C in the inset of figure 2(c).

\begin{figure*}[t]
    \centering
    \includegraphics[width=0.9\textwidth]{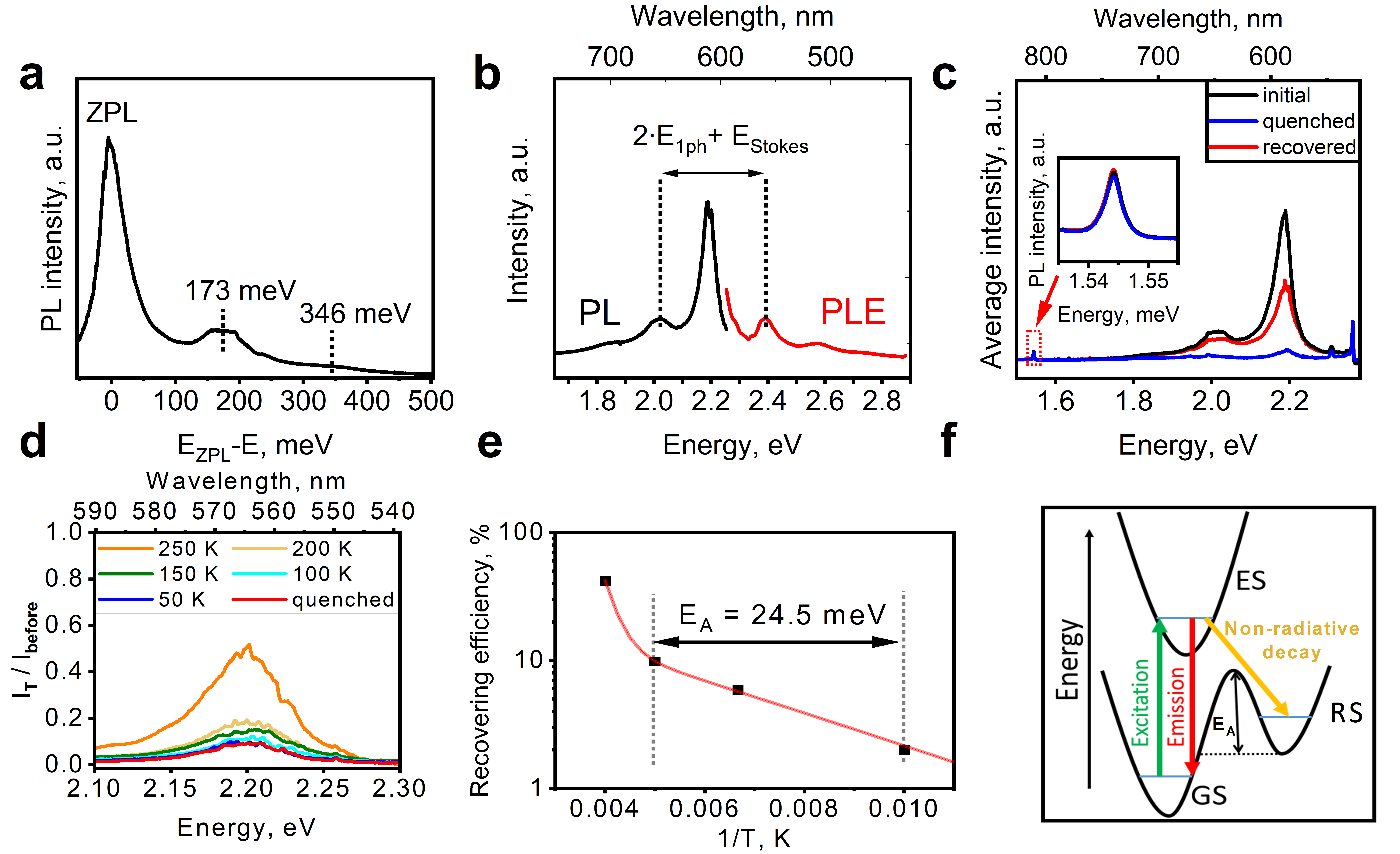}
    \caption{The coupling of the surface defect to phonons and the characterization of the reversible photoluminescence (PL) quenching mechanism. (a) PL spectrum of the surface defect A presented in an energy scale relative to the position of the zero-phonon lines. (b) PL (black curve) and PL excitation (PLE, red curve) spectra for the surface defect A. (c) A spatially averaged low temperature (4~K) PL spectrum of defect A measured via a read-out protocol described in the main text in three states of the sample: 1) initial state (black curve), 2) quenched state (blue curve), and 3) recovered state (red curve). (d) The low temperature (4~K) PL spectra measured after 5 thermal cycling procedures with different upper temperature limits. A thermal cycle involved heating the sample to the upper temperature limit indicated in the legend, maintaining the sample at that temperature for 30 minutes, and cooling it back to 4~K. (e) The dependence of the recovering efficiency (defined in the main text) on the temperature presented in a logarithmic scale as a function of $1/T$. Black points represent the experimental data, and the red curve demonstrates the least-squares fit of a biexponential function. (f) A schematic diagram showing the minimal electronic structure of the defect, which accounts for the reversible quenching mechanism.}
    \label{fig2}
\end{figure*}

The efficiency of the recovery process was found to depend on the upper temperature limit in the thermal cycling process. Higher temperature leads to a systematic increase in the recovery of PL intensity with respect to the initial state, as demonstrated in figure 2(d). The recovering efficiency ($RE$) was quantified as:

\begin{equation}
RE=\frac{I_{initial}-I_{recovered}}{I_{initial}+I_{recovered}}
\end{equation}

where $I_{initial}$ is the spatially averaged total integrated PL intensity of defect A in the initial state of the sample, and $I_{recovered}$ is the spatially averaged total integrated PL intensity after quenching and recovery protocols. The laser power, exposure time, and mapping range were identical for the initial and final read-out of the total PL intensity to ensure comparability.

The dependence of $RE$ on the temperature ($T$) as a function of $1/T$ is demonstrated in figure 2(e). In the low temperature range up to 200 K, we observed a monotonic exponential increase of $RE$, which was reproduced by a thermal activation function:

\begin{equation}
    RE(T)= A\cdot e^{-\frac{E_A}{k_BT}}
\end{equation}

\begin{figure*}[!t]
    \centering
    \includegraphics[width=0.9\textwidth]{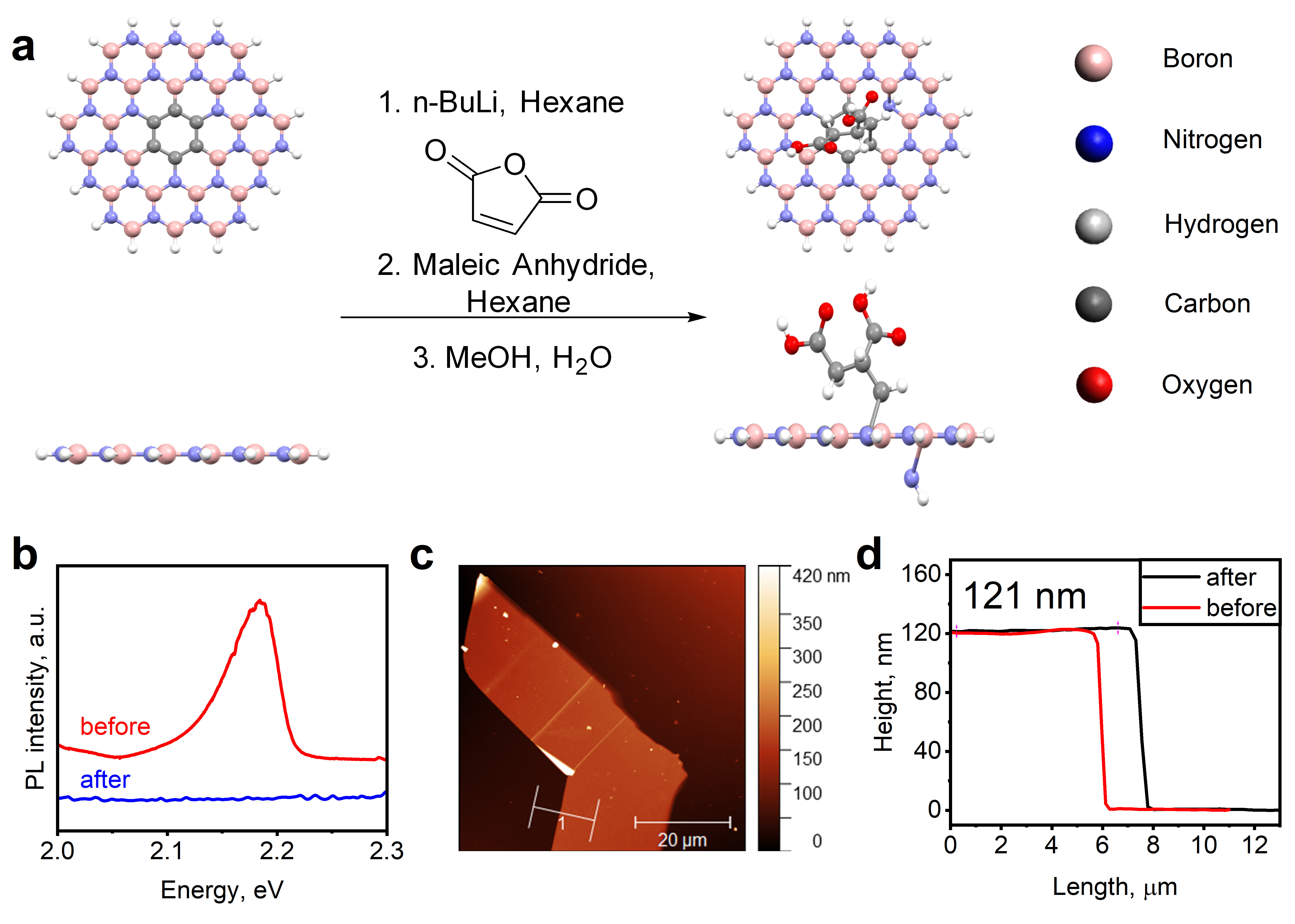}
    \caption{The permanent quenching of the photoluminescence (PL) via surface classical chemistry. (a) The schematic diagram presenting the chemical reactions performed for the carbon-doped hBN flakes and the plausible product of the reactions. (b) Spatially averaged low temperature (4~K) PL spectra of the defect A before (red curve) and after (blue curve) the chemical reactions. (c) Atomic force microscopy image of the carbon-doped hBN flake. (d) The cross-sectional height profiles along the line 1 indicated in panel (c), before (red curve) and after (black curve) the chemical reactions. The two curves are shifted horizontally by 2 $\upmu$m for clarity.}
    \label{fig3}
\end{figure*}

where A is a temperature-independent constant, $E_A$ is the activation energy, and $k_B$ is the Boltzmann constant. The least squares fitting revealed the activation energy  $E_A = 24.5$~meV. An acceleration of the RE rate was observed at higher temperatures (above 200 K), leading to superexponential temperature dependence.

The minimal electronic model that accounts for the quenching and recovery effects is presented in figure 2(f). An optically active transition between a ground state (GS) and an excited state (ES) is responsible for the PL signal observed in the initial state of the sample. The high-power laser illumination provides sufficient energy to induce a metastable reconstructed state (RS), separated by an energy barrier $E_A$ from the GS and enabling an alternative non-radiative decay pathway. The recovery of the PL signal associated with the transition between GS and ES via thermal cycling in a vacuum environment indicates that the RS is related to a structural reconfiguration of the defect rather than a modification of its chemical structure. A plausible mechanism behind such a structural reconfiguration is based on the photoinduced charge transfer between polarized sites within carbon clusters and adjacent defects. 

Such interpretation is further corroborated by the quenching efficiency dependence on the excitation wavelength, shown in figure S1 in the Supplementary Information. We have observed a significantly enhanced quenching efficiency when excitating the defect A resonantly into the optical resonances observed in the PL and PLE spectra, as verified by performing the quenching protocols with 515, 640, 660, 725, and 980 nm lasers under the identical excitation power and exposure time. This underscores the fundamental role of the intradefect optical transition in the formation of the RS. We have also analysed the dynamics of the quenching process. We found that the PL intensity of defect A is decreasing nonlinearly via a biexponential quenching transient with the characteristic quenching time in a few-second range, as illustrated in figure S2 in the Supplementary Information. The quenching dynamics occur in the timescale multiple orders of magnitude longer than the excited state lifetime of the emitter in the ns range (see figure S3 in the Supplementary Information for the measurements of the PL decay transients). The biexponential temporal quenching decay and bimodal temperature evolution of the quenching efficiency imply the existence of multiple possible RS states, with the thermal cycling in the low temperature regime (below 200 K) and fast time decay governed predominantly by the RS state with the lowest $E_A$ of 24.5 meV as extracted from the experimental data.

\textbf{Irreversible quenching via surface chemistry.} 

\begin{figure*}[t]
    \centering
    \includegraphics[width=0.9\textwidth]{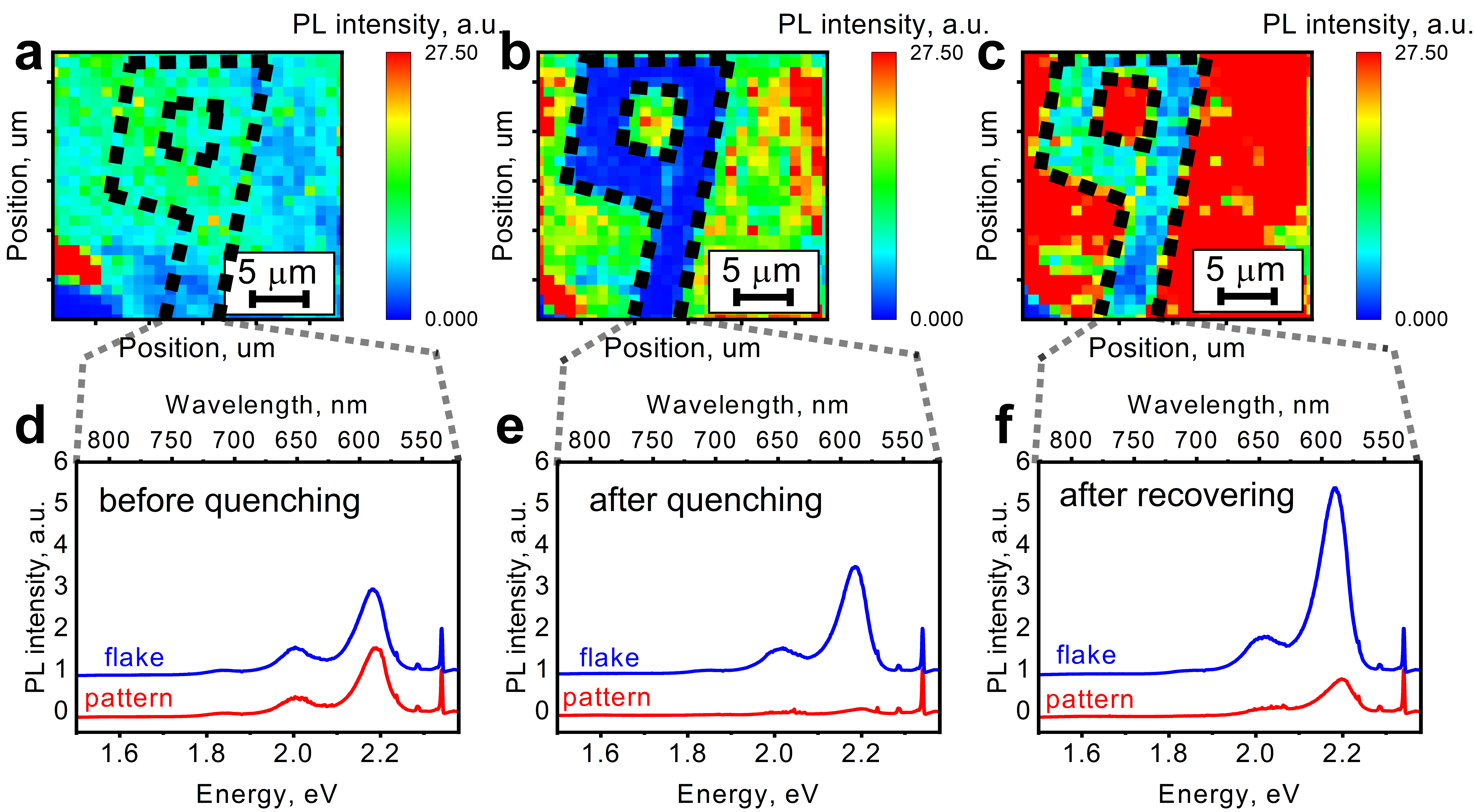}
    \caption{Direct laser writing of negative-contrast emissive patterns. (a-c) Low temperature (4~K) photoluminescence (PL) maps demonstrating the integrated intensity of defect A averaged within and outside of a geometrically defined pattern indicated by a dashed line. The maps were obtained via a read-out protocol described in the main text for three conditions of the sample: 1) initial state (panel a), 2) quenched state achieved via selective irradiation of the patterned area (panel b), and 3) recovered state after a thermal cycling procedure up to room temperature (panel c). (d-f) PL spectra spatially averaged within the pattern area (red curves) and outside of the pattern area (blue curves) for the  (d) initial, (e) quenched, and (f) recovered states of the sample.}
    \label{fig4}
\end{figure*}

An alternative route to permanently quench the PL intensity from defect A was achieved via surface chemistry. We conducted reactions of exfoliated carbon-doped hBN films with n-Butyllithium (lithiation), followed by reaction with electrophilic allylic moiety (maleic anhydride), and oxidative quenching (MeOH, H$_2$O). The reaction with n-Butyllithium occurs through lithiation and charging of carbon fragments, involving $\sigma$+ sites of C or B associated with carbon clusters and the surrounding lattice. The alkyl group involved in lithiation (n-Butyl) does not react with carbon-doped hBN, but the previously charged lithiated sites capped with hydrogen and hydroxyl groups become dislocated from the two-dimensional plane.

Maleic anhydride possesses electron-withdrawing groups coupled to $\pi$ systems.  Therefore, the reaction occurs through the electrophilic addition pathway. The coupling reaction can lead to attachments directly to the carbon cluster or to an adjacent atom neighbouring the negatively charged lithiated carbon inclusion. Furthermore, such reactions are unlikely to result in a quaternary N due to the allylic nature of reactive sites. The most likely scenario involves a carbon cluster becoming activated towards electrophilic addition concurrently at multiple sites.

The reaction pathway and its plausible product are presented in the schematic diagram in figure 3(a). These reactions irreversibly quenched the PL intensity from an ensemble of defects A, as illustrated in figure 3(b). The reactants do not intercalate the hBN interlayer space due to the molecular dimensions significantly exceeding the size of the van der Waals gap, as further evidenced by the unchanged thickness of the hBN films after the reactions (figure 3(c,d)). Therefore, the permanent and complete quenching of defects A via the classical chemistry reactions unambiguously reveals their surface nature. The bulk defects B and C were found to remain unaffected by the surface modifications, as demonstrated in the figure S5 in the Supplementary Information.

The C$_6$ cluster embedded in hBN is more reactive than a freestanding carbon-based aromatic molecule due to $\sigma$+ C polarisation induced by B and $\sigma$- C polarisation induced by N, both on respective ortho and para C sites \cite{baldridge1993balancing}. Therefore, C ipso to N is $\sigma$+ polarized and C ipso to N is $\sigma$- polarized.  As a result, the described reactions will occur on polarized C by modifying the aromatic C$_6$ cluster, consequently disrupting the two-dimensional structure and aromaticity around the reaction site. While adjacent C, B, and N dislocated atoms are typically capped with H or OH groups, the B atoms adjacent to C clusters exhibit higher Lewis acidity due to lower stabilization from C than N in an hBN lattice. In addition, surface B atoms possess higher Lewis acidity due to decreased stabilization from the top layer overlapping N, leading to higher reactivity of such sites.

\textbf{Direct laser writing.}


The unique photophysical characteristics of surface defects A enable the development of resettable negative-contrast direct laser writing. We have designed an arbitrary geometric pattern on a flake showing a finite PL intensity over an entire area of carbon-doped hBN film. The pattern is illustrated in a dashed line in the PL intensity map in figure 4(a) conducted via a read-out protocol at low temperature (4~K). The area of the pattern was selectively irradiated via a quenching protocol, which created a negative-contrast emissive pattern, visualized via a subsequent read-out protocol, as demonstrated in figure 4(b). The recovery protocol via thermal cycling up to room temperature decreased the contrast between the quenched and unquenched regions, as shown in figure 4(c). This represents a signature of partial recovery via thermal cycling, as the activation energy is comparable to the thermal energy at room temperature. Full recovery should be achievable by extending the thermal cycling protocol beyond room temperature. The quantitative comparison of the laser writing and recovery efficiency is provided by the intensity of the  PL signal spatially averaged within ($I_{within}$) and outside ($I_{outside}$) of the pattern, as illustrated in figures 4(d-f) corresponding to the initial, quenched, and recovered states of the sample. The contrast can be defined as a ratio $R=I_{within}/I_{outside}$. The values achieved in the direct laser writing process were: $R_{initial} = 0.95$, $R_{quenched} = 0.05$, and $R_{recovered} = 0.2$.  

\textbf{Conclusion.} 

We demonstrated a novel class of radiative defects in bulk carbon-doped hBN films, which exist at the surface of mechanically exfoliated crystals. The surface defects were found to undergo a laser-induced quenching of the photoluminescence signal, reversible via a thermal cycling process. The mechanism was associated with the formation of a metastable reconstructed state characterized by an activation energy of 24.5 meV for a transition between radiative and non-radiative states. Permanent quenching was achieved via classical surface chemistry reactions, modifying the chemical composition of carbon-related defects. These characteristics were utilized to demonstrate direct laser writing capabilities, developing negative contrast geometric emissive patterns in a microscopic configuration. The existence of surface defects opens further possibilities to explore functionalities in the domain of sensing and optoelectronics by taking advantage of direct surface coupling capabilities.

\vspace{-2.5 mm}

\section*{Acknowledgements}
This project was supported by the Ministry of Education (Singapore) through the Research Centre of Excellence program (grant EDUN C-33-18-279-V12, I-FIM). This research is supported by the Ministry of Education, Singapore, under its Academic Research Fund Tier 2 (MOE-T2EP50122-0012). This material is based upon work supported by the Air Force Office of Scientific Research and the Office of Naval Research Global under award number FA8655-21–1-7026. M. K. acknowledges funding from the MAT-GDT Program at A*STAR via the AME Programmatic Fund by the Agency for Science, Technology and Research under Grant No. M24N4b0034. K.W. and T.T. acknowledge support from the JSPS KAKENHI (Grant Numbers 21H05233 and 23H02052), the CREST (JPMJCR24A5), JST, and World Premier International Research Center Initiative (WPI), MEXT, Japan.

\section*{Author contribution}

M.K. supervised the project.
T.T. and K.W. grew carbon-doped hBN crystals. K.V. fabricated the samples. D.L. and M.G. performed optical measurements. D.L. performed AFM measurements. V.G. and D.L. performed chemical reactions. D.L. and V.G wrote the manuscript. M.K. set the final version of the manuscript. All authors discussed the results.

\section*{Competing interests}
The authors declare no competing interests.

\section*{Methods}
\textbf{Crystal growth.} Single crystals of hBN were grown by the temperature gradient method under high-pressure and high-temperature conditions. To achieve C-doping, selected crystals were annealed for 1-2 h at 2,000 °C with nitrogen gas flow by using a high-frequency furnace with a graphite susceptor. The details of the growth process are discussed in Ref. \cite{hBN_carbon_doping}.

\textbf{Atomic force microscopy.} 
Topography measurements were performed using a Bruker Dimension FastScan XR atomic force microscope operated in tapping and peak force modes with RTESPA-300 cantilevers.

\textbf{Optical spectroscopy.} The optical measurements have been performed in a dry cryogenic system with a base temperature of 4 K. The sample was mounted onto the x-y-z piezo stages. The measurements were realized in a back-scattering geometry, and the laser focusing and collection were done with the cold objective with a numerical aperture of 0.82. For the detection of the photoluminescence signal, the light was dispersed by a SpectraPro HRS-750 spectrometer and detected by a PyLoN 100-BRX LN2-cooled CCD camera (both Princeton Instruments). Photoluminescence excitation measurements were done by using a broadband supercontinuum laser (SuperK Extreme) coupled to the tunable bandpass filter (LLTF contrast).\\ 

\textbf{Chemical reactions.} Carbon-doped hBN flakes on Si/SiO$_2$ substrate were loaded into 100 ml round-bottom flask. The setup was dried with heating (70 °C) under vacuum ($10^{-1}$ mbar) for 3 h, followed by 3 Ar-vacuum cycles. Under an Ar atmosphere, 2 mL 2M n-BuLi in cyclohexane dissolved in 10 mL anhydrous hexane was added. The sample was reacted at 60 °C for 20 h. Maleic anhydride (120 mg) was dried under vacuum with low heat (70~°C), followed by 3 Ar-vacuum cycles. Under an Ar atmosphere, it was dissolved in 40 mL anhydrous hexane, added to the Li-intercalated substrate, and reacted at 70 °C (temperature at external probe) for 20 h. The sample was cooled to room temperature, and the supernatant was cannula extracted. The sample was washed with hexane 2x50 mL, and methanol 4x50 mL (all high-performance liquid chromatography grade).

\clearpage
\newpage
\bibliography{references}

\clearpage
\newpage
\onecolumngrid
\setcounter{figure}{0}
\renewcommand{\thefigure}{S\arabic{figure}}

\section{Supplementary Information}

\textbf{Excitation dependence of the quenching efficiency.}

\vspace{1 mm}

\begin{figure*}[!h]
\centering
\includegraphics[width=0.5\linewidth]{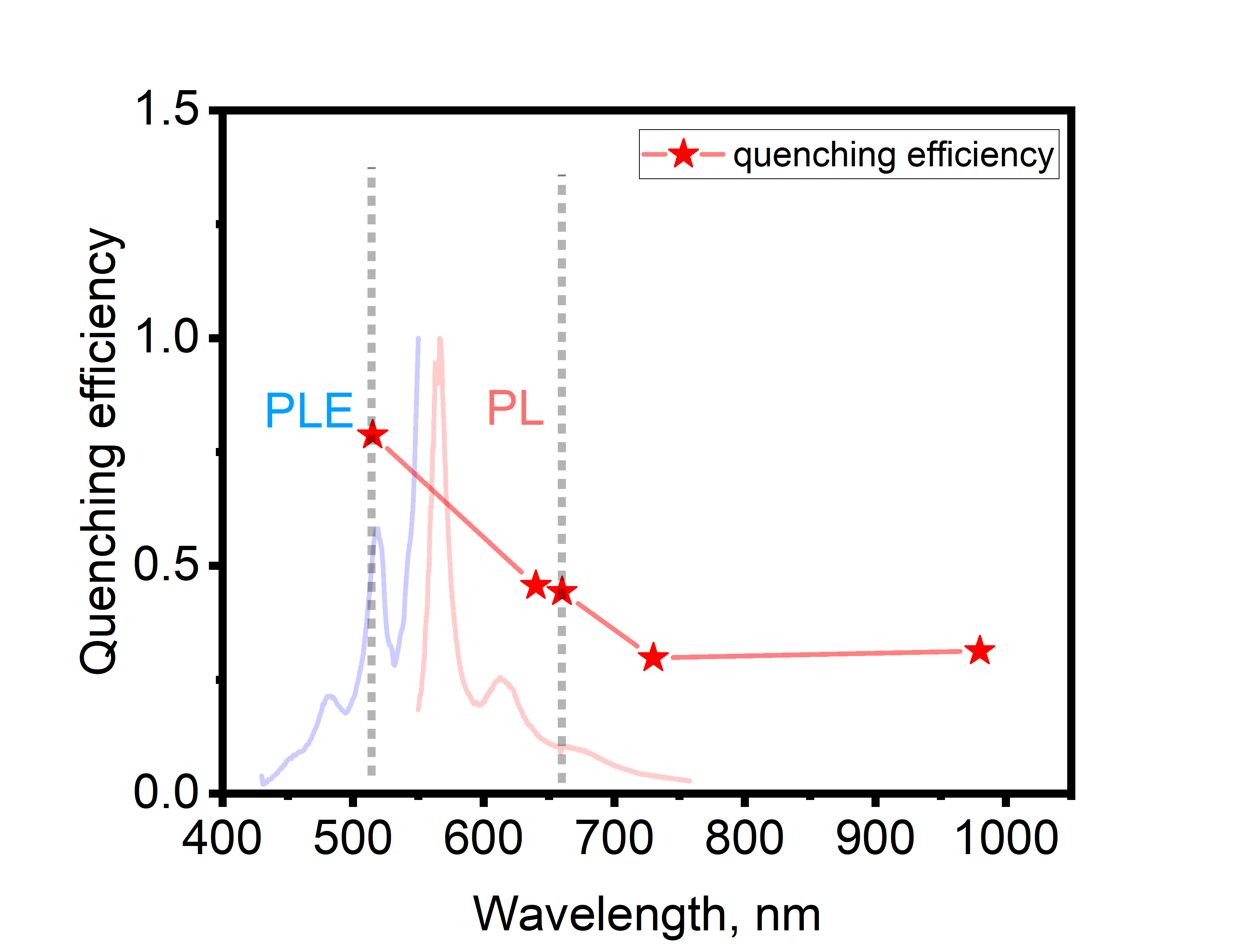}
\caption{Quenching efficiency as a function of excitation wavelength. Blue and red solid curves are photoluminescence excitation and photoluminescence spectra, respectively.}
\label{S2}
\end{figure*}

To gain further insights into the quenching mechanism, we have performed an excitation dependence study on the efficiency of this process. By fixing the power and irradiation time, we exposed the carbon-doped hBN flake to lasers of different wavelengths: 515~nm, 640~nm, 660~nm, 725~nm, and 980~nm. By defining the quenching efficiency as $1-\frac{I_{quenched}}{I_{initial}}$, figure \ref{S2} shows the quenching efficiency as a function of the laser wavelength. As can be seen, the highest quenching has been observed with 515 nm, when the laser energy matches a quasi-absorption feature of the defect (the blue curve demonstrates the photoluminescence excitation spectrum). 

The quenching efficiency decreases with increasing wavelength up to 725 nm. This result demonstrates that optical excitation of the defect constitutes the dominant mechanism leading to the transition to the dark state.\\

\textbf{Quenching dynamics.}

\vspace{1 mm}

\begin{figure*}[!h]
\centering
\includegraphics[width=0.8\linewidth]{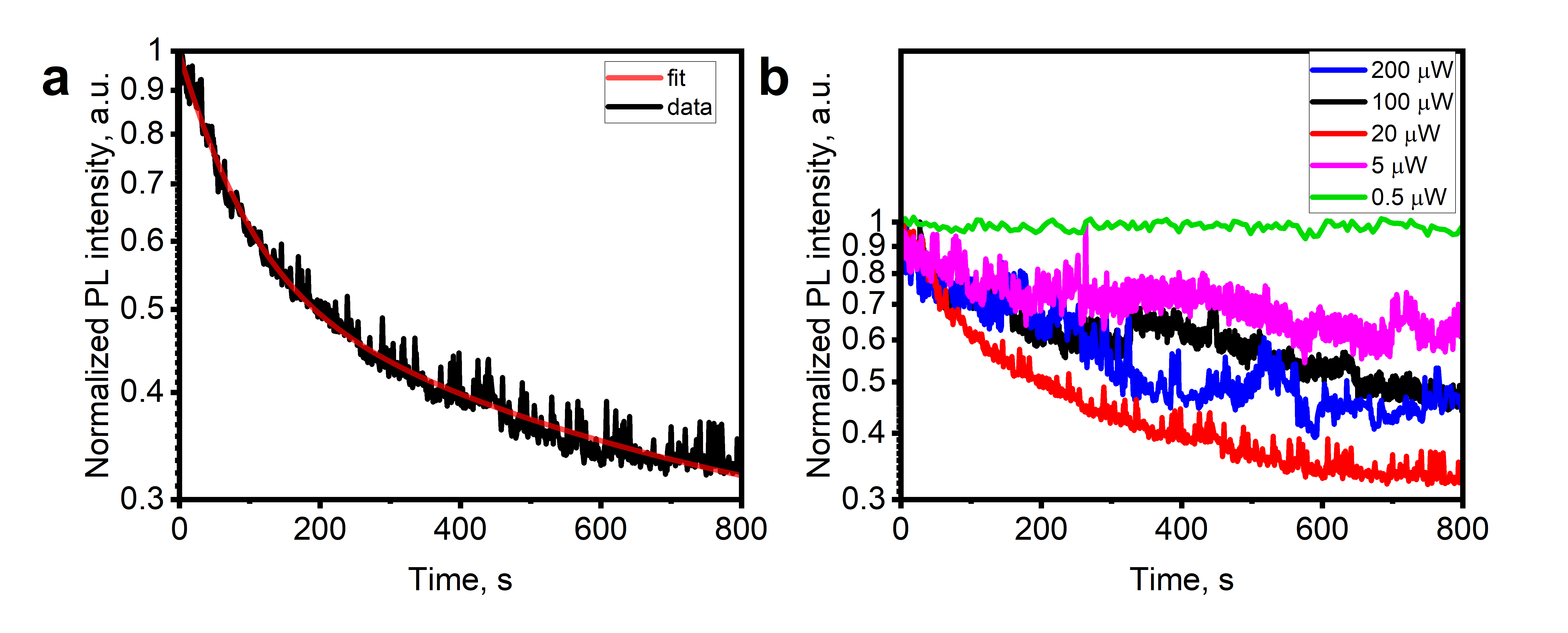}
\caption{(a) Normalized integrated photoluminescence intensity of the defect A fitted by a biexponential decay function measured for 20 $\upmu$W laser excitation. (b) Normalized integrated photoluminescence intensity of the defect A as a function of time recorded under different excitation powers.}
\label{S1}
\end{figure*}

To analyze the dynamics of the quenching process, we have performed time-dependent photoluminescence (PL) intensity measurements. Figure~\ref{S1}(a) presents the normalized integrated PL intensities as a function of time. 

The best fit has been obtained by using the biexponential decay model:

\begin{equation}
    I=y_o+A_1\cdot e^-{\frac{x}{T_1}}+A_2\cdot e^-{\frac{x}{T_2}}
\end{equation}

Where $T_1$ and $T_2$ are related to the transition rate to different reconstructed states. Figure~\ref{S1}(b) presents the power-dependent decay measurements. As can be seen, at the low power (500 nW), there is almost no quenching presented. The rate is increasing with the power up to 20 $\upmu$W, which is expected for the structural transition discussed in the main text. At further increase in the power, we observed the reduction of the quenching rate (blue and red curves in the figure~\ref{S2}(b). This may be related to the local heating, which is forcing the thermal relaxation of defect A back to the ground state.\\

\textbf{Photoluminescence decay transient.}

\vspace{1 mm}

\begin{figure}[!h]
    \centering
    \includegraphics[width=.8\linewidth]{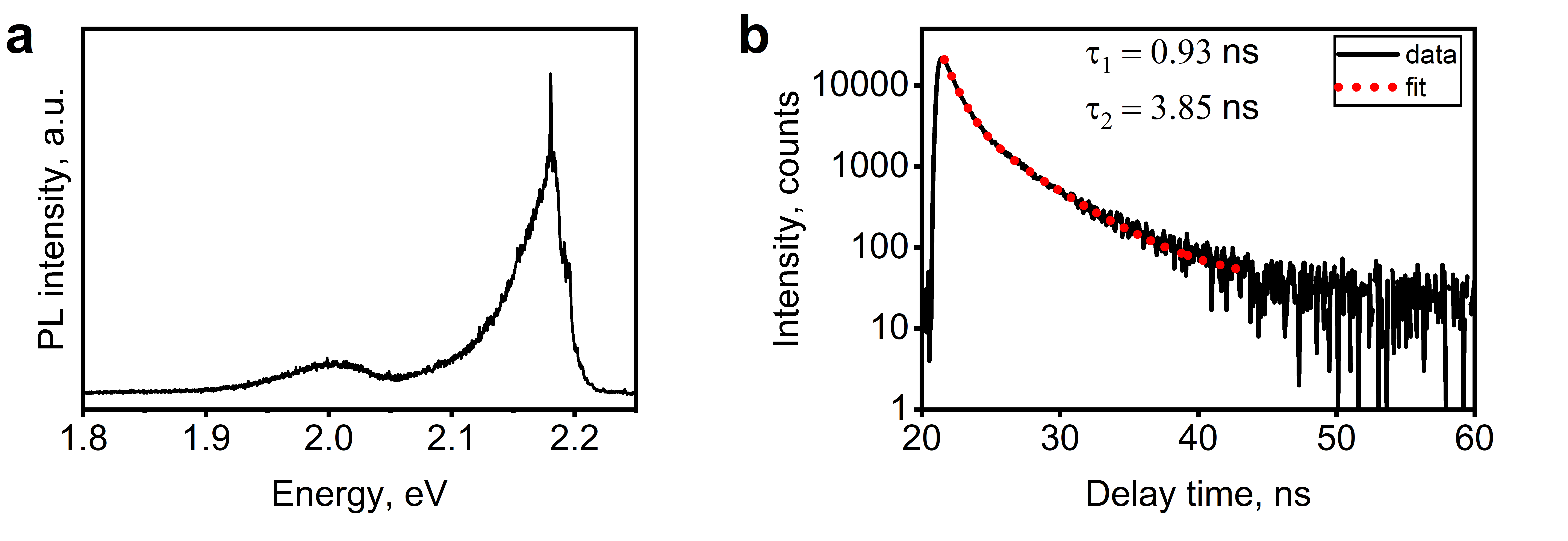}
\caption{a) Photoluminescence spectrum of the defect A. b) The photoluminescence decay transient of an ensemble of defects A.}
    \label{fig.rep1}
\end{figure}

The time-resolved PL measurements and the PL spectrum obtained at the investigated microscopic location of the carbon-doped hBN flake are presented in Figure~\ref{fig.rep1}.

The PL spectrum (figure~\ref{fig.rep1} a) demonstrates the typical spectral position of the zero-phonon line for the emitter A discussed in the main text with the pronounced phonon replica. The best fit of the time decay transient (figure~S\ref{fig.rep1} b)  has been obtained with the biexponential function:

\begin{equation}
    I(t)=y_0+A_1e^{-\frac{x}{\tau_1}}+A_2e^{-\frac{x}{\tau_2}}
    \label{eq1}
\end{equation}

where $\tau_1$ and $\tau_2$ are the lifetimes of the corresponding states and have been found to be 0.93 ns and 3.85 ns, respectively. These values are typical for different types of radiative defects in hBN.\\

\textbf{Spatial resolution of the negative-contrast laser writing.}

\vspace{1 mm}

The diffraction-limited beam size for our system (NA = 0.82 and wavelength of 515 nm) used in the quenching protocol is 766 nm. The intensity profile of the laser beam is given as:

\begin{equation}
    I(r)=\frac{2 P}{\pi\omega}e^{-\frac{2r^2}{\omega^2}}
    \label{eq 2}
\end{equation}

\noindent
where r is the distance from the center of the beam, $\omega$ is the diameter of the beam, and P is the laser power.

\begin{figure}[!h]
    \centering
    \includegraphics[width=.9\linewidth]{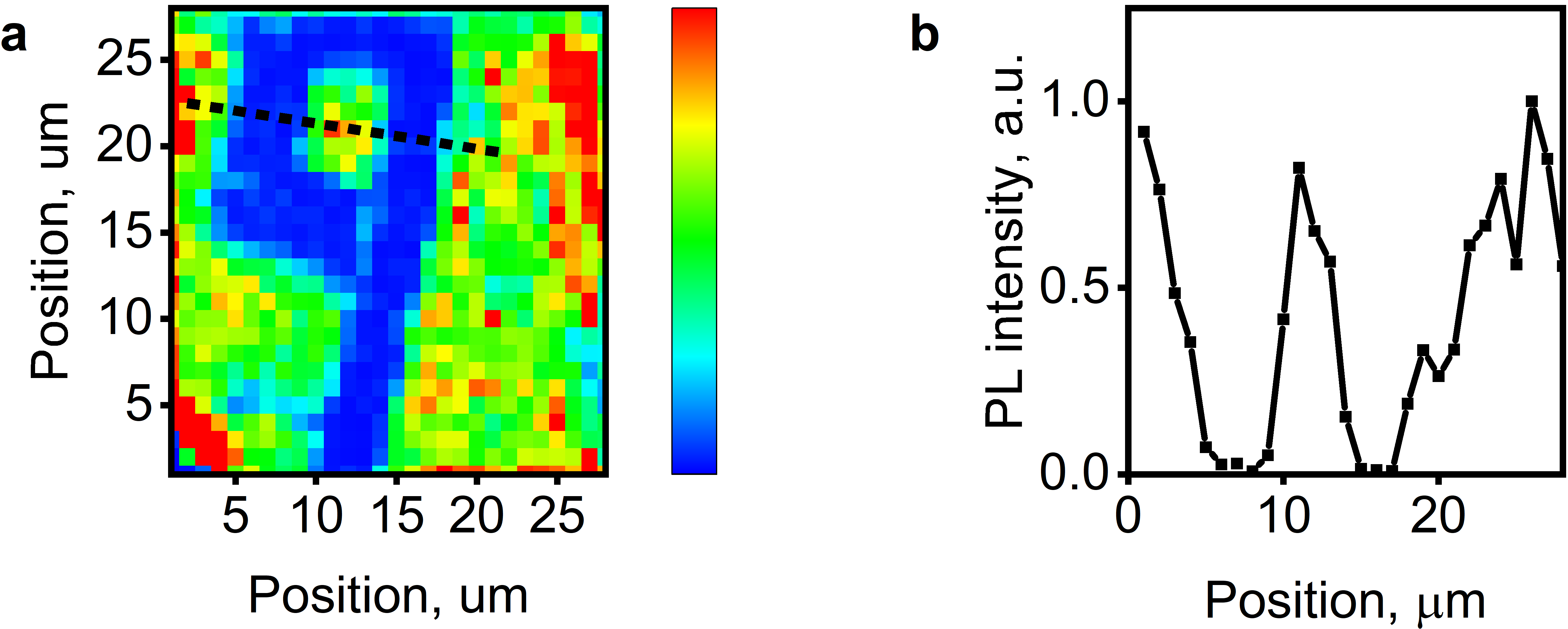}
\caption{Spatial resolution of the negative laser writing technique based on the deterministic quenching of defect emission. a) The map of the integrated photoluminescence intensity of the defect A after quenching. b) The cross-section of the map presented in (a) along the dashed black line.}
    \label{fig.rep2}
\end{figure}

For our patterning experiment, we used the laser power of 10 mW, which should lead to a quenching area of 2.7 $\mu$m  diameter based on the estimation of the onset of quenching power density. The cross-sectional profile of our experimentally achieved quenched pattern shows the minimal feature size of about 3 $\mu$m (see Fig.~\ref{fig.rep2}), which agrees well with the estimation achieved by using the equation \ref{eq 2}.

It is important to note that the resolution could be improved further. Beyond the optimization of the process parameters (laser wavelength, laser power, exposure time, or sample temperature), it is possible to laser write features smaller than the diffraction limit, e.g., by employing two-photon absorption processes.\\

\textbf{Quenching of defects via surface chemistry.}

\vspace{1 mm}

\begin{figure}[!h]
    \centering
    \includegraphics[width=.9\linewidth]{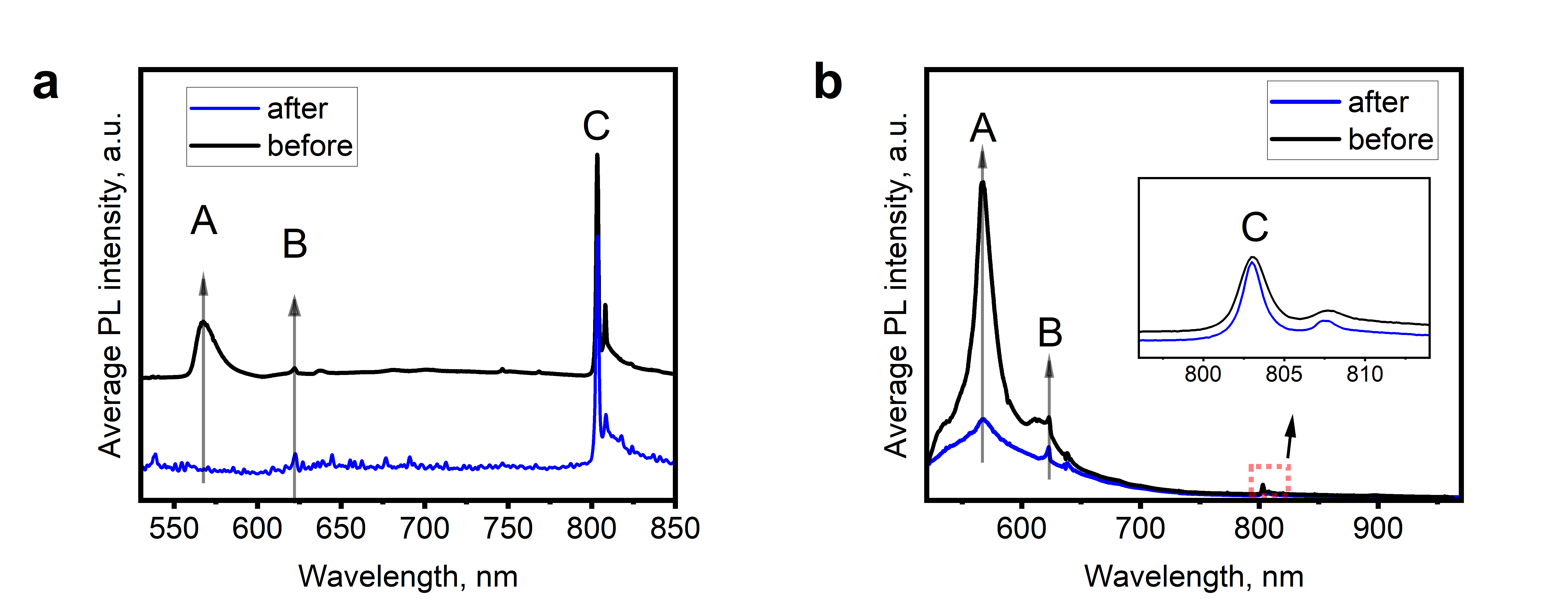}
\caption{Irreversible quenching of surface emitters via chemical functionalization. Average PL spectrum obtained by the integration of the PL intensity over the flake area from the hyperspectral map before (black curves) and after (blue curves) the chemical reaction: a) lithiation followed by reaction with maleic anhydride, and b) lithiation with the oxidative quenching.}
    \label{fig.rep3}
\end{figure}

We have performed two different reactions to validate the surface termination of defects A: 1) lithiation followed by an electrofilic moiety, and 2) lithiation with the oxidative quenching. Corresponding PL spectra before and after reactions are presented in Figure \ref{fig.rep3}. For both reactions, we have observed a significant reduction of the PL intensity for defects A, while the emission characteristics for bulk defects B and C remained unaffected. 

\end{document}